\documentclass[aps,prl,twocolumn,showpacs]{revtex4}

\newcommand{\nc}{\newcommand}

\nc{\gtwid}{\mathrel{\raise.3ex\hbox{$>$\kern-.75em\lower1ex\hbox{$\sim$}}}}
\nc{\ltwid}{\mathrel{\raise.3ex\hbox{$<$\kern-.75em\lower1ex\hbox{$\sim$}}}}
\nc{\comp}{{\rm C}\llap{\vrule height7.1pt width1pt depth-.4pt\phantom t}}
\nc{\Dal}{\kern1pt\vbox{\hrule height 1.2pt\hbox{\vrule width 1.2pt\hskip 3pt
   \vbox{\vskip 6pt}\hskip 3pt\vrule width 0.6pt}\hrule height 0.6pt}\kern1pt}

\begin{document}

\preprint{UFIFT-HEP-03-05,astro-ph/0307358}

\title{A generic problem with purely metric formulations of MOND}

\author{M. E. Soussa}
\email[]{soussam@phys.ufl.edu}

\author{R. P. Woodard}
\email[]{woodard@phys.ufl.edu}
\affiliation{Department of Physics, University of Florida,
             Gainesville, FL 32611, USA}

\date{\today}

\begin{abstract}

We give a simple argument to show that no purely metric-based, relativistic 
formulation of Milgrom's Modified Newtonian Dynamics (MOND) whose energy
functional is stable (in the sense of being quadratic in perturbations) can 
be consistent with the observed amount of gravitational lensing from galaxies. 
An important part of the argument is the fact that reproducing the MOND force 
law requires any completely stable, metric-based theory of gravity to become 
conformally invariant in the weak field limit. We discuss the prospects for 
a formulation with a very weak instability.

\end{abstract}

\pacs{04.50.+h, 11.10.Lm, 98.80.Cq}

\maketitle

{\it 1. Introduction:} A large spiral galaxy might contain a mass of
$M \sim 10^{11} M_{\odot} \sim 10^{41}~{\rm kg}$ in the form of stars
and gas. Almost all of this mass lies inside a radius of $R \sim 10~{\rm
kpc} \sim 10^{20\!-\!21}~{\rm m}$. With numbers on these scales it is easy
to see that the gravitational acceleration ought to be Newtonian,
\begin{equation}
\frac{G M}{R^2} \sim 10^{-10}\, \frac{\rm m}{{\rm ~s}^2} \; .
\label{azero}
\end{equation}
Hence neutral Hydrogen in a circular orbit of radius $r > R$ should
be observed to move more slowly as $r$ increases,
\begin{equation}
\frac{v^2_N}{r} = \frac{G M}{r^2} \qquad \Longrightarrow \qquad 
v^2_N(r) = \frac{G M}{r} \; . \label{Newt}
\end{equation}
As is well known, it does not. In the scores of galaxies for which high 
quality rotation curves can be obtained \cite{SanMc}, one actually finds that 
the asymptotic speed approaches a constant which is proportional to the 
fourth root of the total luminosity. This is known as the Tully-Fisher 
relation \cite{TF}.

Dark matter is the conventional explanation for the observed failure of the 
Newtonian prediction (\ref{Newt}). It was originally invoked to provide the
mass necessary for Newtonian gravitation to bind galaxies to their host
clusters \cite{Zwicky,Smith}. To explain flat rotation curves it is supposed 
that galaxies are surrounded by halos of weakly-interacting, non-relativistic 
particles whose contribution to the total mass density falls off much more 
slowly than the mass density in stars and gas \cite{KK,Turner}.

Dark matter has been impressively successful in reconciling the observed
universe with general relativity and its Newtonian limit \cite{Turner}.
However, it is not without problems. First, there has been no direct detection
of the particles which comprise it, except for neutrinos which can provide
only a small fraction of the necessary mass and which are too light to cluster
on galaxy scales. Second, what accounts for galactic rotation curves is an 
isothermal halo $\rho_{\rm iso} \sim 1/r^2$ of dark matter. But numerical 
simulations of dark matter consistently produce the density profile of 
Navarro, Frenk and White \cite{NFW}, which falls like $\rho_{\rm NFW} \sim 
1/r^3$ at large distances. Dark matter does not explain the Tully-Fisher
relation. Nor can it elucidate the curious fact that the breakdown of 
Newtonian gravity, sourced by only the mass in stars and gas, seems to 
occur at the same characteristic acceleration (\ref{azero}) in cosmic 
structures of vastly different sizes. (But see \cite{KapTur}.)

Milgrom has proposed that, instead of dark matter, galactic rotation curves 
signal a modification of gravity at very low accelerations \cite{Milgrom}. If 
Newtonian theory predicts a gravitational acceleration $\vec{a}_{\rm N}$, 
then the gravitational acceleration in Milgrom's theory (MOND) is,
\begin{equation}
\vec{a}_{\rm M} = f\Bigl(\frac{a_{\rm N}}{a_0}\Bigr) \vec{a}_{\rm N}
\quad {\rm where} \quad f(x) = \cases{ 1 & $\forall x \gg 1$ , \cr
x^{\!-\! \frac12} & $\forall x \ll 1$ .}
\end{equation}
The MOND interpolating function $f(x)$ is assumed to vary smoothly between the
stated limits. Many forms are consistent with the data. A typical example is,
\begin{equation}
f(x) = \sqrt{\frac12 + \frac12 \sqrt{1 + \frac4{x^2}}} \; . \label{exf}
\end{equation}
In MOND $a_0$ has the status of a new universal constant. Its numerical value 
has been determined by fitting to nine well-measured galaxies 
\cite{BBS},
\begin{equation}
a_0 = (1.20 \pm .27) \times 10^{-10} \, \frac{\rm m}{{\rm ~s}^2} \; .
\end{equation}

MOND was designed to explain the Tully-Fisher relation. In the MOND regime
of $a_N/a_0 \ll 1$ one finds,
\begin{equation}
\frac{v^2}{r} \longrightarrow \frac{\sqrt{a_0 G M}}{r} \quad \Longrightarrow
\quad v^2 \longrightarrow v^2_{\infty} \equiv \sqrt{a_0 G M} \; . \label{TFE}
\end{equation}
Making the natural assumption that the total luminosity $L$ is proportional 
to the total mass one finds, $v^4_{\infty} = a_0 G M \propto L$, which is the
Tully-Fisher relation \cite{TF}. 

MOND not only reproduces the asymptotic rotation velocity $v_{\infty}$, it 
also describes the inner portions of rotation curves. MOND fits have been
worked up for on the order of 100 galaxies, using only the mass-to-luminosity
ratios in stars and in gas as free parameters. The recent review paper by 
Sanders and McGaugh \cite{SanMc} summarizes the data and lists the primary 
sources. Except for a handful of galaxies that show evidence of recent 
disturbance, the fits are excellent. Further, the mass-to-luminosity ratios 
obtained from the fits are in rough agreement with evolution models \cite{BJ}. 
It is especially significant that MOND even fits the rotation curves of low 
surface brightness galaxies \cite{MB,BM}, structures for which the MOND 
regime of $a_N/a_0 \ltwid 1$ applies throughout and for which no detailed 
measurements had been made when the theory was proposed. Although some dark 
matter is needed to explain the temperature and density profiles of large 
galaxy clusters \cite{ASQ}, it has been suggested that this might be provided 
by neutrinos without affecting the galaxy results \cite{Sanders1}.

The chief problem with MOND is that it does not provide a complete theory
of gravitation the way that general relativity does. Bekenstein and Milgrom
\cite{BekMil} have given a satisfactory field theory for the nonrelativistic
potential whose negative gradient gives the gravitational acceleration,
$\vec{a} = -\vec{\nabla} \phi$. If $\rho_m$ is the mass density, the
Lagrangian for $\phi$ is,
\begin{equation}
{\cal L} = - \rho_m \phi - \frac{a_0^2}{8 \pi G} F\Bigl(\frac{\Vert 
\vec{\nabla} \phi \Vert^2}{a_0^2}\Bigr) \; ,
\end{equation}
where the interpolating function associated with (\ref{exf}) would be
$F(x) = \sqrt{x + x^2} - \ln(\sqrt{x} + \sqrt{1 + x})$. The existence of
this Lagrangian is very important because it establishes that MOND 
conserves energy, momentum and angular momentum. However, this model
provides no information about the other gravitational potentials, or about 
time evolution. One must therefore make essentially {\it ad hoc} assumptions 
to test what MOND says about lensing \cite{MorTurn1}, or about cosmology 
\cite{Sanders2}. 

What is needed is a generally covariant formulation of MOND that includes
at least the usual metric. A generally coordinate invariant, scalar-metric 
extension of the Bekenstein-Milgrom theory exists \cite{BekMil} but its 
prediction for the deflection of star light is identical to that of general 
relativity. Without dark matter, this gives far too little gravitational
lensing from galaxies to be consistent with the data \cite{MorTurn2}. 
Scalar-metric models which reproduce MOND and may be consistent with 
extra-galactic lensing data can be constructed, but only at the price of 
allowing the scalar to carry negative energy \cite{Bek,BekSan}.

Earlier this year we devised a generally covariant, purely metric model of 
MOND based on a nonlocal effective action \cite{SW}. Although we were able 
to reproduce the MOND force law, our model also failed to give the extra
deflection of star light that would be provided in general relativity by dark 
matter. After studying the problem we have concluded that it is generic: any 
stable, generally covariant and metric-based theory of gravity that 
reproduces the MOND force law must suffer the same problem of too little 
lensing. The purpose of this letter is to give a careful presentation of the 
argument. We do this in section 2. Our model is presented in section 3 to
illustrate the problem. Section 4 identifies the five assumptions that
lead to the generic problem. We also discuss the prospects for abandoning 
one of them.

{\it 2. The argument:} We assume that the gravitational force is carried by 
the metric tensor $g_{\mu\nu}(x)$ and that its source is the usual 
stress-energy tensor $T_{\mu\nu}(x)$. This implies that the metric is 
determined by a set of ten field equations having the form,
\begin{equation}
{\cal E}_{\mu\nu}[g] = \frac{8 \pi G}{c^4} T_{\mu\nu} \; . \label{genform}
\end{equation}
We also assume that the theory of gravitation is generally covariant. This
implies conservation,
\begin{equation}
g^{\rho \sigma} {\cal E}_{\mu \rho ; \sigma} = 0 \; , \label{cons}
\end{equation}
where the semi-colin denotes covariant differentiation. In general relativity
${\cal E}_{\mu\nu}[g]$ would be the Einstein tensor, $G_{\mu\nu}$, but we
make no assumptions about ${\cal E}_{\mu\nu}$ at this stage. In particular, 
we allow it to involve higher derivatives, and even to be a nonlocal 
functional of the metric.

Now recall that deviations between MOND and Newtonian gravitation become
significant only for very small accelerations. These accelerations arise 
from de\-riv\-a\-tives of the metric. General coordinate invariance allows us 
to choose a coordinate system in which the metric agrees with the Minkowski
metric $\eta_{\mu\nu}$ at a single point. If its gradients are also small
--- which is the observed fact --- then the metric can be made numerically
quite close to $\eta_{\mu\nu}$ over a large region. It therefore suffices to 
study equations (\ref{genform}) using weak field perturbation theory around 
flat space. That is, we write,
\begin{equation}
g_{\mu\nu}(x) = \eta_{\mu\nu} + h_{\mu\nu}(x) \; ,
\end{equation}
and we assume the numerical magnitude of the graviton field $h_{\mu\nu}(x)$ 
is small in the MOND regime. 

We wish to consider the weak field expansion of ${\cal E}_{\mu\nu}[\eta + h]$.
In general relativity this begins at linear order, but that cannot be the
case for any theory which reproduces the MOND force law. To see why, consider
the gravitational response to a static distribution of total mass $M$, such
as a low surface brightness galaxy, whose density is low enough that the 
equations (\ref{genform}) are everywhere in the MOND regime. The MOND 
explanation (\ref{TFE}) for the Tully-Fisher relation implies that at least 
one component of $h_{\mu\nu}$ must scale like $\sqrt{G M}$. (For spherical 
distributions it would be the $rr$ component, but this does not matter.)
How can (\ref{genform}) result in such a dependence? Since the right-hand 
side scales like $G M$, it follows that at least one tensor component of 
${\cal E}_{\mu\nu}[\eta + h]$ must go like $h^2$.

If we assume gravity is absolutely stable then all ten components of 
${\cal E}_{\mu\nu}[\eta + h]$ cannot begin at quadratic order in the
weak field expansion. This is because the dynamical subset of the field
equations follow from varying the gravitational Hamiltonian. If its
variation were quadratic then the Hamiltonian would be cubic, and this 
is not consistent with stability. We therefore conclude that only a
subset of the ten components of ${\cal E}_{\mu\nu}[ \eta + h]$ can go
like $h^2$.

The desired subset must be distinguished in some generally covariant fashion.
A symmetric, second rank tensor field contains two distinguished subsets: its
divergence and its trace. The weak field expansion of the divergence does
vanish a linearized order, but we see from conservation (\ref{cons}) that it
vanishes to all orders. So the divergence cannot be responsible for the 
required $h^2$ term and we are left with the trace as the only remaining
possibility,
\begin{equation}
g^{\mu\nu} {\cal E}_{\mu\nu}[\eta + h] = O(h^2) \; . \label{notrace}
\end{equation}

Equation (\ref{notrace}) can explain the Tully-Fisher relation (\ref{TFE}),
but it means that MOND corrections to general relativity can be removed,
in the weak field limit, by a local conformal rescaling of the metric,
\begin{equation}
g_{\mu\nu}(x) \longrightarrow \Omega^2(x) g_{\mu\nu}(x) \; . \label{conftrans}
\end{equation}
To see why, recall that traceless metric field equations imply invariance 
under the transformation (\ref{conftrans}). (This has been known for decades. 
For a proof, see \cite{TW}.) The full field equations are not traceless, so 
neither is the full theory conformally invariant. However, because the 
linearized field equations are traceless, the linearized theory must be
conformally invariant. This means that the linearized field equations only 
determine the metric up to a conformal factor (and, of course, up to a 
linearized diffeomorphism). The conformal part of the metric is fixed by the 
order $h^2$ term in the trace of the field equations, and this is how one 
component of the weak field $h_{\mu\nu}$ contrives to go like $\sqrt{G M}$. 
But these MOND corrections to general relativity must be removable by a 
conformal rescaling (\ref{conftrans}), and possibly a simultaneous coordinate 
transformation.

This is a disaster for the phenomenology of gravitational lensing. To see why
recall that, for a general metric $g_{\mu\nu}$, the Lagrangian of 
electromagnetism is,
\begin{equation}
{\cal L} = -\frac14 F_{\mu\nu} F_{\rho \sigma} g^{\mu\rho} g^{\nu\sigma}
\sqrt{-g} \; ,
\end{equation}
where $F_{\mu\nu} \equiv \partial_{\mu} A_{\nu} - \partial_{\nu} A_{\mu}$.
This Lagrangian is invariant under a conformal rescaling (\ref{conftrans}),
which means that electromagnetism is unaware of MOND corrections to general
relativity in the weak field limit. Hence the deflection of star light will
be only that predicted by general relativity. Without dark matter this gives
far too little lensing from galaxies \cite{MorTurn2}. That is the generic
problem mentioned in the title of this paper.

{\it 3. An example:} Although the argument we have given is completely
general, it is illuminating to see how the problem arises in a specific
model. For this purpose we review the model whose failure motivated this
study \cite{SW}. It is based on two nonlocal functionals of the metric.
The first of these is known as the {\it small potential},
\begin{equation}
\varphi[g] \equiv \frac{1}{\Dal} R \;\; {\rm where} \;\; \Dal \equiv
\frac{1}{\sqrt{-g}} \partial_{\mu} \Bigl( \sqrt{-g} g^{\mu\nu} \partial_{\nu}
\Bigr) \; . \label{smallpot}
\end{equation}
We use a spacelike metric with $R_{\mu\nu} \equiv \Gamma^{\rho}_{~ \nu\mu , 
\rho} \!-\! \Gamma^{\rho}_{~ \rho \mu , \nu} \!+\! \Gamma^{\rho}_{~ \rho 
\sigma} \Gamma^{\sigma}_{~ \nu \mu} \!-\! \Gamma^{\rho}_{~ \nu \sigma} 
\Gamma^{\sigma}_{~ \rho \mu}$. The covariant d'Alembertian is inverted with 
retarded boundary conditions so that $\varphi[g](x)$ depends only upon the 
metric and its derivatives evaluated on or inside the past light-cone of 
the point $x^{\mu}$. 

The second nonlocal functional in our construction is called, the {\it large 
potential},
\begin{equation}
\Phi[g] \equiv \frac{1}{\Dal} \Bigl[ g^{\rho\sigma} \varphi_{,\rho} 
{\cal F}'\!\Bigl(c^4 a_0^{\!-\!2} g^{\mu\nu} \varphi_{,\mu} \varphi_{,\nu}
\Bigr) \Bigr]_{;\sigma} \; . \label{largepot}
\end{equation}
To reproduce the MOND force law it turns out that ${\cal F}(x)$ must obey
\cite{SW},
\begin{equation}
{\cal F}(x) = - \frac{x}{6} - \frac{x^{\frac32}}{162} + O(x^2) \; .
\end{equation}

For our model the functional ${\cal E}_{\mu\nu}[g]$ is \cite{SW},
\begin{eqnarray}
\lefteqn{{\cal E}_{\mu\nu}[g] \!=\!  2 [ \Phi_{;\mu\nu} \!-\! g_{\mu\nu} \Dal 
\Phi] \!+\! G_{\mu \nu} [1 \!-\! 2 \Phi] } \nonumber \\
& & \!\!\!\!\!+\! g_{\mu\nu} \varphi^{,\rho} \Phi_{,\rho} \!-\! \varphi_{,\mu} 
\Phi_{,\nu} \!-\! \varphi_{,\nu} \Phi_{,\mu} \!+\! \varphi_{,\mu} 
\varphi_{, \nu} {\cal F}' \!-\! \frac{a_0^2}{2 c^4} g_{\mu\nu} {\cal F} . 
\qquad \label{MONDeqn}
\end{eqnarray}
This form is manifestly covariant. It can be checked that it is also
conserved. Although ${\cal E}_{\mu\nu}[g](x)$ is nonlocal in our model, 
it is causal in the sense of depending only upon the fields on or inside the
past light-cone of the point $x^{\mu}$.

To prove asymptotic conformal invariance, note first that although the weak
field expansion of the Ricci tensor $R_{\mu\nu}$ involves arbitrary powers of 
the weak fields, it begins at linear order. This implies that the weak field
expansion of the small potential (\ref{smallpot}) also begins at linear order.
The same applies to the large potential (\ref{largepot}) because it is
proportional to the small potential in the MOND limit of ${\cal F}(x) \!
\rightarrow \! -\frac16 x$,
\begin{equation}
\Phi[g] = \frac{1}{\Dal} \Bigl( \varphi^{,\rho} {\cal F}' \Bigr)_{;\rho} 
\longrightarrow -\frac16 \varphi \; .
\end{equation}
In taking the weak field limit of ${\cal E}_{\mu\nu}$ we can therefore neglect 
any products of $R_{\mu\nu}$, $\varphi$ or $\Phi$, such as $G_{\mu\nu} \Phi$, 
$\varphi^{,\rho} \Phi_{,\rho}$ and $\varphi_{,\mu} \varphi_{,\nu}$. We can 
also neglect ${\cal F} \sim \varphi^{,\rho} \varphi_{,\rho}$. Hence the weak
field limit of ${\cal E}_{\mu\nu}$ is contained in just four terms,
\begin{equation}
{\cal E}_{\mu\nu} \longrightarrow \frac13 \Bigl(g_{\mu\nu} \Dal \varphi - 
\varphi_{;\mu\nu} \Bigr) + R_{\mu\nu} -\frac12 g_{\mu\nu} R \; .
\end{equation}
Of course these terms involve many higher powers of $h_{\mu\nu}$, but they
contain {\it all} the linear powers. And the terms we have kept are exactly
traceless,
\begin{equation}
g^{\mu\nu} \Bigl(\frac13 g_{\mu\nu} \Dal \varphi - \frac13 
\varphi_{;\mu\nu} + R_{\mu\nu} - \frac12 g_{\mu\nu} R \Bigr) = \Dal 
\varphi - R = 0 .
\end{equation}

Note that tracelessness (and hence conformal invariance) is not a property of 
the full theory. In general the trace gives,
\begin{eqnarray}
\lefteqn{g^{\mu\nu} {\cal E}_{\mu\nu} = -6 \Dal \Phi - R [1 - 2 \Phi]}
\nonumber \\
& & \hspace{1.5cm} + 2 \varphi^{,\mu} \Phi_{,\mu} + \varphi^{,\mu} 
\varphi_{,\mu} {\cal F}' - \frac{2 a_0^2}{c^4} {\cal F} \; .
\end{eqnarray}
This goes like $h^2$ in the weak field expansion. It can be shown that the
MOND terms in the force law derive entirely from the resulting quadratic 
equation \cite{SW}, in accord with the general argument of section 2.

{\it 4. Discussion:} The point of making a no-go argument is to identify
the assumptions which result in the negative conclusion. It can then be 
considered which, if any, of the assumptions might be discarded. For the 
argument given above we made the following assumptions:
\begin{itemize}
\item The gravitational force is carried by the metric, and its source is
the usual stress-energy tensor.
\item The theory of gravitation is generally covariant.
\item The MOND force law is realized in weak field perturbation theory.
\item The theory of gravitation is absolutely stable.
\item Electromagnetism couples conformally to gravity.
\end{itemize}

It seems to us that the weakest assumption is absolute stability. This is what 
dictated that only a subset of the ten components of ${\cal E}_{\mu\nu}[g]$ 
can be quadratic in the weak field expansion. If we abandon absolute stability
it becomes possible that {\it all ten} components of ${\cal E}_{\mu\nu}[g]$
are quadratic in the weak fields. 

This may not be as bad as it sounds. It should be understood that any theory 
of MOND necessarily possesses two weak field regimes: the ultra-weak limit 
in which MOND applies, and a less-weak regime in which gravity is weak 
but MOND corrections to general relativity are negligible. It is the latter 
regime which describes the solar system and the interior of our galaxy, so 
gravity would still be stable in these settings. 

The instability could only manifest in regions of very low gravitational 
acceleration. Even there it might be self-limiting because the creation of 
any significant density of decay products --- whatever they are --- would 
likely drive the theory back into the less-weak regime in which it is stable. 
So one might imagine a universe that very gradually decays, in the empty 
regions between galaxies, into long wave length particles whose density is 
diffused as the universe expands. If $a_0$ is really a constant this decay
would only have started recently in cosmological history. And its further
progress must be heavily suppressed by the intrinsic weakness of the 
gravitational interaction. We therefore conclude it is worth searching for 
a generally covariant, metric-based formulation of MOND in which all ten 
components of the field equations are quadratic in the weak field limit. 

Even if no viable, generally covariant formulation of MOND can be constructed,
this would in no way invalidate the impressive observational data that has 
been accumulated over many years \cite{SanMc}. The absence of an acceptable 
generalization of MOND would mean that this data is not explained by an
alternate theory of gravitation, but we wish to stress that {\it the data
must still be explained}. Either the low acceleration regime of gravity is
ruled by some generalization of MOND or else isolated distributions of dark 
matter evolve towards some hitherto unrecognized attractor solution. Both
alternatives are fascinating, and we feel the community is greatly indebted
to those whose patient labors have drawn attention to the problem.

{\it 5. Acknowledgements:} It is a pleasure to acknowledge discussions and
correspondence with J. D. Bekenstein, M. Milgrom and R. H. Sanders. This work 
was partially supported by DOE contract DE-FG02-97ER41029 and by the 
Institute for Fundamental Theory.

\vskip -.5cm
%%%%%%%%%%%%%%%%%%%%%%%%%%%%%%%%%%%%%%%%%%%%%%%%%%%%%%%%%%%%%%%%%%%%%%%%%%%%%%%
%    REFERENCES
%%%%%%%%%%%%%%%%%%%%%%%%%%%%%%%%%%%%%%%%%%%%%%%%%%%%%%%%%%%%%%%%%%%%%%%%%%%%%%%


\begin{thebibliography}{99}
\bibliographystyle{unsrt}

\bibitem{SanMc} R. H. Sanders and S. S. McGaugh, A. Rev. Astron. \& Astrophys.
{\bf 40}, 263 (2002), arXiv: astro-ph/0204521.

\bibitem{TF} R. B. Tully and J. R. Fisher, Astron. Astrophys. {\bf 54}, 661
(1977).

\bibitem{Zwicky} F. Zwicky, Helvetica Physica Acta {\bf 6}, 110 (1933).

\bibitem{Smith} S. Smith, Astrophys. J. {\bf 83}, 23 (1936).

\bibitem{KK} G. R. Knapp and J. Kormendy, {\it Dark Matter in the Universe}
(IAU Symposium \#117; Reidel, Dordrecht, 1987).

\bibitem{Turner} M. S. Turner, Phys. Rept. {\bf 333}, 619 (2000).

\bibitem{NFW} J. F. Navarro, C. S. Frenk and S. D. M. White, Astrophys. J.
{\bf 490}, 493 (1997).

\bibitem{KapTur} M. Kaplinghat and M. Turner, Astrophys. J. {\bf 569}, L19
(2002), arXiv: astro-ph/0107284.

\bibitem{Milgrom} M. Milgrom, Astrophys. J. {\bf 270}, 365 (1983).

\bibitem{BBS} K. G. Begeman, A. H. Broeils and R. H. Sanders, MNRAS {\bf 240},
523 (1991).

\bibitem{BJ} E. F. Bell and R. S. de Jong, Astrophys. J. {\bf 550}, 212
(2001), arXiv: astro-ph/0011493.

\bibitem{MB} S. S. McGaugh and W. J. G. de Blok, Astrophys. J. {\bf 449}, 66
(1998), arXiv: astro-ph/9801102.

\bibitem{BM} W. J. G. de Blok and S. S. McGaugh, Astrophys. J. {\bf 508}, 132
(1998), arXiv: astro-ph/9805120.

\bibitem{ASQ} A. Aguirre, J. Schaye and E. Quataert, Astrophys. J. {\bf 561},
550 (2001), arXiv: astro-phys/0105184.

\bibitem{Sanders1} R. H. Sanders, MNRAS {\bf 342}, 901 (2003), arXiv: 
astro-ph/0212293.

\bibitem{BekMil} J. D. Bekenstein and M. Milgrom, Astrophys. J. {\bf 286}, 7
(1984).

\bibitem{MorTurn1} D. J. Mortlock and E. L. Turner, Mon. Not. R. Astron. Soc.
{\bf 372}, 557 (2001), arXiv: astro-ph/0106100.

\bibitem{Sanders2} R. H. Sanders, Astrophys. J. {\bf 560}, 1 (2001), arXiv:
astro-ph/0011439.

\bibitem{MorTurn2} D. J. Mortlock and E. L. Turner, Mon. Not. R. Astron. Soc.
{\bf 372}, 552 (2001), arXiv: astro-ph/0106099.

\bibitem{Bek} J. D. Bekenstein, Phys. Rev. {\bf D48}, 3641 (1993), arXiv:
gr-qc/9211017.
 
\bibitem{BekSan} J. D. Bekenstein and R. H. Sanders, Astrophys. J. {\bf 
429}, 480 (19994), arXiv: astro-ph/9311062.

\bibitem{SW} M. E. Soussa and R. P. Woodard, Class. Quant. Grav. {\bf 20},
2737 (2003), arXiv: astro-ph/0302030.

\bibitem{TW} N. C. Tsamis and R. P. Woodard, Ann. Phys. {\bf 168}, 457 (1986).

\end{thebibliography}
\end{document}